\documentclass[%
 reprint,
 amsmath,amssymb,
 aps,
floatfix,
]{revtex4-2}

\usepackage{graphicx}
\usepackage{dcolumn}
\usepackage{bm}

\usepackage[x11names]{xcolor}
\usepackage{framed}
\usepackage{quoting}

\usepackage{appendix}
\usepackage{lipsum}

\definecolor{beaublue}{rgb}{0.74, 0.83, 0.9}
\definecolor{cerulean}{rgb}{0.0, 0.48, 0.65}
\definecolor{frenchblue}{rgb}{0.0, 0.45, 0.73}

 \colorlet{shadecolor}{beaublue}
\newenvironment{shadedquotation}
 {\begin{shaded*}
  \quoting[leftmargin=0pt, vskip=0pt]
 }
 {\endquoting
 \end{shaded*}
}

\begin{document}

\preprint{APS/123-QED}

\title{Individual risk-aversion responses tune epidemics \\ to critical transmissibility ($R=1$)}

\author{Susanna Manrubia}
\affiliation{Department of Systems Biology, National Centre for Biotechnology (CSIC). \\ c/ Darwin 3, 28049 Madrid, Spain \\
Interdisciplinary Group of Complex Systems (GISC), Madrid, Spain}%


\author{Damián H. Zanette}
\affiliation{
 Centro Atómico Bariloche  and Instituto Balseiro,\\
 Comisión Nacional de Energía Atómica and Universidad Nacional de Cuyo \\
 Consejo Nacional de Investigaciones Científicas y Técnicas \\
 Av.~E.~Bustillo 9500, 8400 San Carlos de Bariloche, Río Negro, Argentina
}%


\date{\today}

\begin{abstract}
Changes in human behavior are increasingly recognized as a major determinant of epidemic dynamics. Although collective activity can be modified through imposed measures to control epidemic progression, spontaneous changes can also arise as a result of uncoordinated individual responses to the perceived risk of contagion. Here we introduce a stochastic epidemic model that implements population responses driven by individual- and time-dependent risk-taking propensity. The model reveals an emergent mechanism for the generation of multiple infection waves of decreasing amplitude without the need to consider external modulation of parameters. Successive waves tune the effective reproduction number to its critical value $R=1$. This process is a consequence of the interplay of the fractions of susceptible and infected population and the average risk-taking propensity, as shown by a mean-field approach. The proposed mechanism also shows how, under the threat of contagion, the distribution of individual risk propensities evolves towards a well-defined profile. Successive waves trigger selective sweeps of risk-taking propensity at a pace determined by individual risk reaction rates. This kind of collective, self-generated pressure, may therefore shape risk-aversion profiles associated to epidemics in human groups. The final state is self-organized and generic, independent of the parameter values. We conclude that uncoordinated changes in human behavior can, by themselves, explain major qualitative and quantitative features of the epidemic process, as the emergence of multiple waves and the tendency to remain around $R=1$ observed worldwide after the first few waves of COVID-19. 
\end{abstract}

\maketitle


\begin{shadedquotation}
\noindent  
\textcolor{frenchblue}
{{\bf Significance statement} \\
Individual perception of external risk usually triggers protective responses. In epidemics, current levels of incidence or the number of deaths caused by the disease prompts individual reactions towards effectively decreasing the number of contacts with potentially infected individuals. In turn, decreases in risk perception and the overcoming of risky situations increase individual confidence and relaxes risk awareness. Here we show that a population of individuals characterized by a stochastic variation of their risk-taking propensity in response to an expanding epidemic can cause multiple epidemic waves that lead the system to the critical value of the epidemic reproduction number $R=1$. The epidemic acts as a filtering process that selects a specific profile for the basal risk-taking propensity of the population.}
\end{shadedquotation}

\section{Introduction} 
 
Self-initiated behavior of individuals aware of an external risk is a major determinant of the course of an epidemic outbreak \cite{ferguson:2007}. Historical examples abound where epidemic spread has been inhibited without the explicit adoption of imposed institutional measures \cite{funk:2010}. For instance, the final incidence of the Ebola outbreak in West Africa in 2014 was well below the predictions of models including a variety of contention measures \cite{pandey:2014} but missing the effect of uncoordinated changes in public habits \cite{butler:2014}. More recently, detailed models of COVID-19 progression in Illinois \cite{wong:2020} have locally underestimated the number of cases due to the non-compliance of infected individuals with the rule of observing quarantine. However difficult capturing human behavior might be, social responses are main drivers of epidemic dynamics \cite{vanBavel:2020}, with consequences that might overrun those derived from any other intervention. It is unknown to which extent social behavior alone, in the absence of external actions, can limit epidemic damage, prevent epidemic extinction, or generate major epidemic features such as persistent waves. Its accurate, quantitative incorporation into epidemic models \cite{verelst:2016} is an open challenge of the discipline. 

At odds with classical descriptions of epidemic dynamics, dominated by single, isolated waves,  and the concept of herd immunity as the mechanism that leads to outbreak termination, COVID-19 progression displays sub-exponential growth, waves, and long plateaus in incidence \cite{castro:2020,thurner:2020,weitz:2020,tkachenko:2021}. There seems to be a growing consensus towards the dependence of these features on mobility \cite{kraemer:2020,mazzoli:2020} and temporal variations in individual activity \cite{tkachenko:2021}, on a heterogeneous social structure (e.g. superspreaders \cite{nielsen:2021}), and on awareness-driven human actions ---including the application of contention measures and collective changes in human behavior \cite{amaral:2021}. 

Some models have considered different qualitative ways in which changes in human habits responding to an external threat could affect epidemic propagation \cite{epstein:2008,funk:2010,scarpino:2016}. Risk awareness, for example, was introduced to show that local transmission of information might suffice to inhibit an epidemic outbreak in spatial settings \cite{funk:2009}. Short-  and long-term awareness has been considered in various mean-field models of epidemic progression \cite{weitz:2020}, and it has been shown that predictions of accumulated incidence improve under the iterative incorporation of behavioral changes \cite{eksin:2019}. Adaptive behavior and change of contact patterns can also generate infection waves, through mechanisms such as the application and release of containment measures \cite{tkachenko:2021}, incoming fluxes of individuals \cite{juher:2020}, delays in the progression of the disease \cite{weitz:2020,johnston:2020} or in the time of exposure of different individual groups \cite{just:2018}, or external seasonal forcing \cite{tkachenko:2021}. Plateaus observed in the epidemic incidence of COVID-19 have been explained through death-awareness \cite{weitz:2020}, and through individual heterogeneity and stochasticity in social activity levels \cite{tkachenko:2021,tkachenko:2021-PNAS}. 

Changes in behavior occur when contention measures are applied at a regional or country level. These measures are well-defined and their effects can be measured to a certain extent \cite{flaxman:2020}. Spontaneous changes in individual behavior due to subjective evaluation of an external threat are more difficult to quantify. Not surprisingly, individual attitude towards risk is a topic of high interest in psychology. Empirical studies have shown that risk aversion has individual- and  time-dependent components \cite{jung:2015}, while transitory increases might be linked to extreme events \cite{gilson:2015}. There is a significant growth of risk aversion with age \cite{paulsen:2012}, but studies of how exposure to extreme events modifies baseline risk preferences are inconclusive \cite{schildberg:2018}. In the framework of the COVID-19 pandemic,  some research has addressed the relationship between risk perception and individual responses. In agreement with former studies on risk-taking propensity, it has been shown that early awareness of the risk posed by COVID-19 ---and a concomitant increase of protective behavior--- was uneven among individuals \cite{wise:2020}, showing widespread heterogeneity in risk-taking \cite{guenther:2021}. Also, individual age, risk-taking propensity, and concern about the pandemic were good predictors of the adoption of precautionary behavior \cite{thoma:2021}. Personal experience with COVID-19 did not seem to have an impact in the overall risk perception of the disease \cite{attema:2021}, thus pointing at a population-averaged perception of risk.  In agreement with the above, risk perception of COVID-19 around the world positively correlated with the adoption of preventative health behaviors \cite{dryhurst:2020}. 

Here we focus on the effects of feedback between stochastic and uncoordinated responses of risk-aware individuals and an ongoing epidemic, in the absence of {\it ad hoc} mechanisms such as
external modulations or delayed reactions. The relevant variable in our formulation is the momentary risk-taking propensity of each individual. In the light of current knowledge, we model a heterogeneous population of individuals, each characterized by its own basal risk-taking propensity, whose exposure to risk is modified as a function of the epidemic incidence. Our analysis yields two main results: first, unsupervised individual responses cause infection waves and select for values of the epidemic effective reproduction number progressively closer to its critical value $R=1$; second, contagion preferentially affects individuals with higher basal and momentary values of risk-taking propensity, asymptotically shaping the risk-aversion profile of the population. 

\section*{Epidemic model with stochastic individual exposure to external risk}

Risk-taking propensity (RTP) is a complex, subjective, and multifactorial personal attribute. In the present model, we interpret this variable as the frequency of individual exposition to an external threat (contagion, in this case), which results from a combination of voluntary actions and unavoidable facts. We consider a population formed by $N$ individuals. For each of them, we introduce  a time-dependent probability $p_i(t)$ ($i=1,\dots , N$) which controls the frequency at which he/she becomes exposed (consistently, the complementary probability $1-p_i(t)$ measures risk aversion). Each individual, moreover, is assigned a basal RTP $p_i^0$, drawn at random from a prescribed distribution. As described below, this basal RTP is an upper bound for $p_i(t)$.

The evolution of individual RTP is driven by two mechanisms, which act in opposite directions (see Methods). At each time step, $p_i(t)$ decreases in an amount $\omega q(t)$, where $q(t)$ is the fraction of infected population. The coefficient $\omega$ weights how fast RTP is inhibited, and thus quantifies the risk-aversion response to the epidemic. To avoid that $p_i (t)$ reaches zero or negative values, its decrease is limited to a small random level accounting for the minimum risk that individuals are forced to face. 

Growth of $p_i(t)$, leading to relaxation towards its basal value $p_i^0$, occurs every time the individual is exposed to contagion but is not infected. In this case, $p_i(t)$ grows in an amount $\rho \xi (t)$, where $\rho$ controls the speed at which risk aversion subsides, and $\xi (t) \in (0,1)$ is a random number. This increase  stops when $p_i(t)$ reaches $p_i^0$, thus saturating at its basal value.  

\begin{figure}[t]
\centering
\includegraphics[width=\columnwidth]{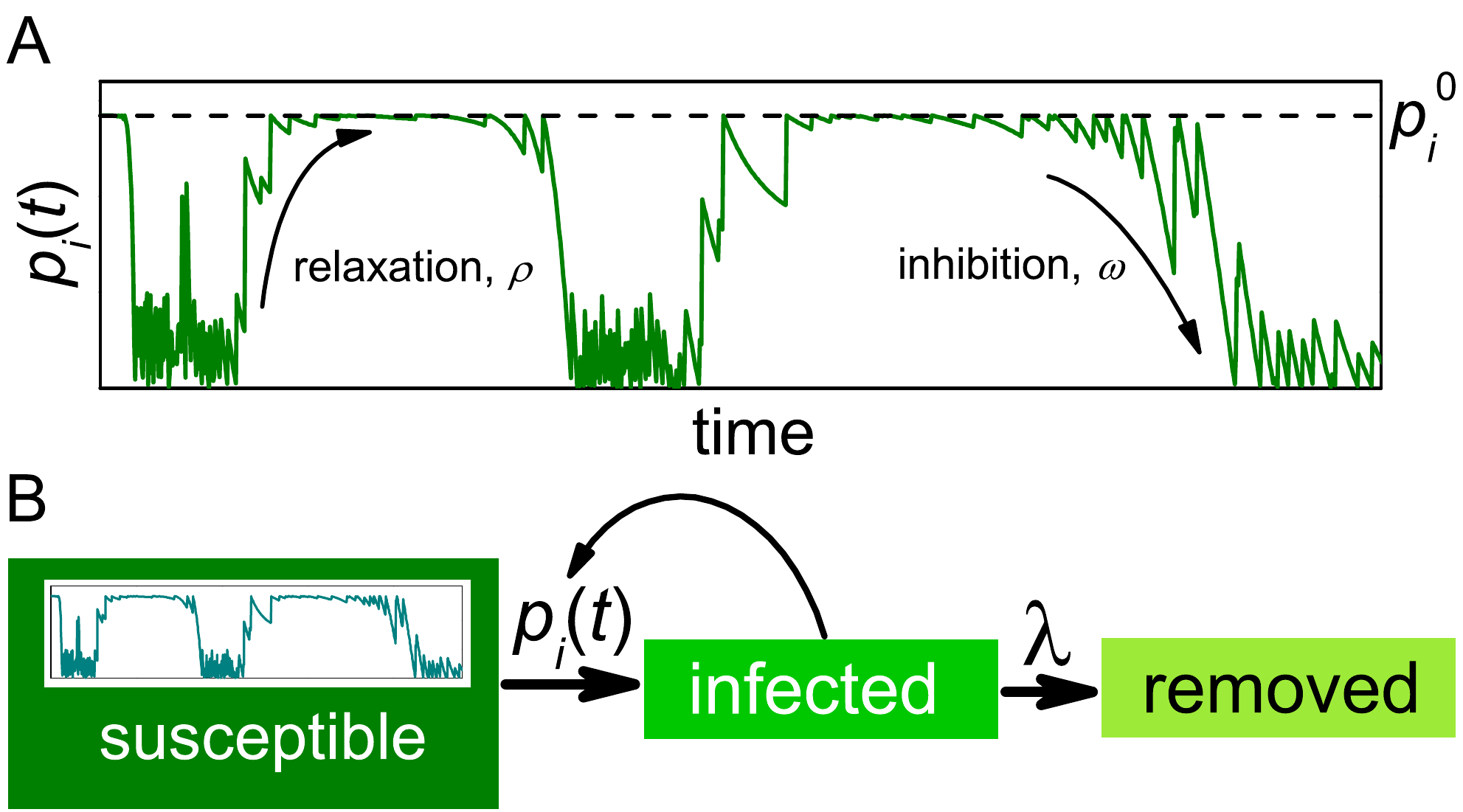}
\caption{Schematics of the  epidemic model. (A) Representative example of the stochastic dynamics of the individual's RTP $p_i(t)$. The dashed horizontal line shows the individual's basal RTP, $p_i^0$. At each time step, individuals inhibit their RTP in an amount proportional to the level of external risk, which is given by the fraction of infected population, weighted by a parameter $\omega$. If, upon exposure to contagion, they do not become infected, their RTP relaxes back to its basal value, controlled by a parameter $\rho$. (B) Exposure to risk of susceptible individuals is stochastic, so that the transition from susceptible to infected occurs with variable probability. Infected individuals are removed with probability $\lambda$.}
\label{fig:SchematicModel}
\end{figure}

Figure~\ref{fig:SchematicModel} illustrates the characteristic dynamics of $p_i(t)$, and the transitions between epidemiological classes, which correspond to those of an SIR model. Individuals can be in states S (susceptible), I (infected) or R (removed). As described in detail in Methods, the stochastic transition from S to I depends on $p_i(t)$, while the transition from I to R occurs with constant probability $\lambda$, as in traditional SIR dynamics. 

For a disease with a typical time span $\lambda^{-1}$, the fraction of infected population $q(t)$ is comparable to the average number of individuals who became infected during an interval of $\lambda^{-1}$ time steps. This latter quantity coincides with the measure that seems to represent better the current state of the COVID-19 pandemic, namely, weekly or biweekly averages of incidence. After considering several other variables (as the number of new cases, excess mortality, number of hospitalizations or fraction of tested population), biweekly averages have emerged as a stable and representative measure, thus supporting our choice to couple individual reactions to $q(t)$. 

\section*{Results}

\subsection*{Individual response inhibits epidemic propagation and induces infection waves}

Under the rules described above, a collective response of the population emerges. We define 
\begin{equation}
    P(t)=\frac{1}{S(t)}\sum_{\{S(t)\}}p_i(t)
\end{equation} 
as the average RTP of the susceptible population at time $t$, where $S(t)$ is the current number of susceptible individuals. Assuming that the basal RTPs $p_i^0$ are uniformly distributed in $(0,1)$, and that initially $p_i(0)=p_i^0$, we have $P(0) \approx 0.5$.  As soon as the number of infected individuals begins to grow, the average RTP decreases because of two reasons: more risk-propense individuals are infected first and, on the average, the epidemic triggers a protective response towards lower values of $p_i(t)$. 

For $\omega=\rho=0$, the individuals do not respond to the progress of the infection, and maintain their basal RTP, regardless of the threat. The epidemic spreads, preferentially affecting individuals with higher $p_i^0$, until it reaches a peak and its incidence declines. A single epidemic wave, comparable to that of a standard SIR model, results. However, when the response of the individuals is turned on ($\omega, \rho > 0$), the generalized decrease in RTP causes a lower peak, while the number of susceptible individuals when the outbreak finishes is higher than without population response. Moreover, the final value of $P(t)$ is higher, meaning that prudent behavior has also protected individuals with higher basal RTP (see Fig. S1). 

\begin{figure}[t]
\centering
\includegraphics[width=.9\columnwidth]{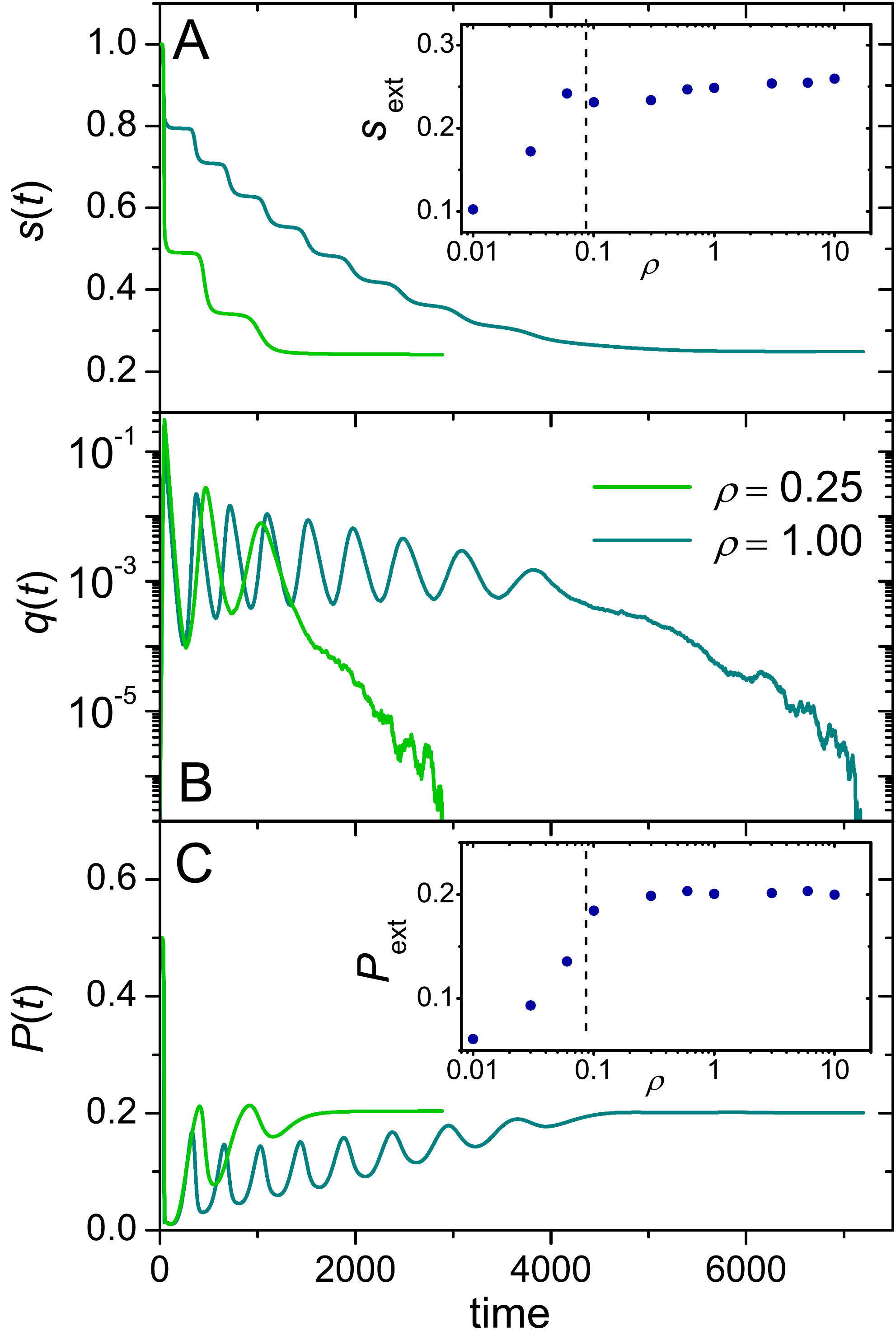}
\caption{Dynamics of the model for a population of $N=10^7$ individuals, with $\lambda=0.05$, $\omega=0.05$, and the two values of $\rho$ indicated in the legend of panel B. (A) Fraction of susceptible individuals, $s(t)$. The inset shows the value of $s(t)$ at the time of extinction, $s_{\rm ext}$, for different values of $\rho$, averaged over ten realizations  for each value. The vertical dashed line indicates the transition from one to two infection waves, $\rho \approx 0.0865$. To the right of this value, two or more waves occur. (B) Fraction of infected individuals, $q(t)$. (C) Average RTP over the susceptible population, $P(t)$. Data in the inset correspond to the value of $P(t)$ at the time of extinction, $P_{\rm ext}$, averaged over the same realizations as in the inset of  panel A.}
\label{fig:ModelDynamics}
\end{figure}

As a function of $\omega$ and $\rho$, one, two or more infection waves can occur. The mechanism behind multiple wave generation can be understood by means of a mean-field approximated representation of the stochastic model, discussed in a later section. Figure \ref{fig:ModelDynamics} illustrates two situations with three ($\omega=0.05$, $\rho=0.25$) and nine waves  ($\omega=0.05$, $\rho=1$). Other parameters being equal, in fact, infection waves become more numerous as $\rho$ grows. The fraction of susceptible individuals decreases through a series of steps  produced by the succession of infection waves. The  amplitude of waves becomes progressively smaller, and their frequency grows monotonically (with exceptional stochastic departures due to finite-population effects). Depending on the parameters, as the number of waves grows, the final decrease of $q(t)$ slows down, and a shoulder or plateau of variable length develops (see Fig.~S2). 

Concomitantly with the infection waves, the average RTP $P(t)$ oscillates in a series of inhibition-relaxation cycles. After an abrupt drop caused by the first wave, $P(t)$ recovers higher values as the remaining susceptible individuals relax their RTP. Superimposed to the oscillations, this recovery occurs over an emerging slow time scale (of the order of a few thousand time steps in Fig. \ref{fig:ModelDynamics}C). Finally,  $P(t)$ stabilizes at a fixed value until all infected individual are removed and the epidemic becomes extinct at a time $t_{\rm ext}$, which grows with $\rho$ (see Fig. S2). Our numerical simulations show that the average RTP and the fraction of susceptible population at the time of extinction, $P_{\rm ext}$ and $s_{\rm ext}$, are weakly dependent on $\omega$ and $\rho$. This is illustrated in the insets of Fig. \ref{fig:ModelDynamics}. 

\subsection*{Multiple infection waves lead to $R=1$}

The effective reproduction number of an epidemic, $R(t)$, is defined as the number of secondary infections caused by each infected individual. In our simulations, $R(t)$ can be calculated as the ratio between the fraction $q_n(t+1)$ of newly infected individuals at time $t+1$ and the fraction of infected population at time $t$, times the average duration of the infection,  given by $\lambda^{-1}$:
\begin{equation}
    R(t) = \frac{1}{\lambda} \frac{q_n(t+1)}{q(t)} \, .
\end{equation}
 
\begin{figure}[t]
\centering
\includegraphics[width=.9\columnwidth]{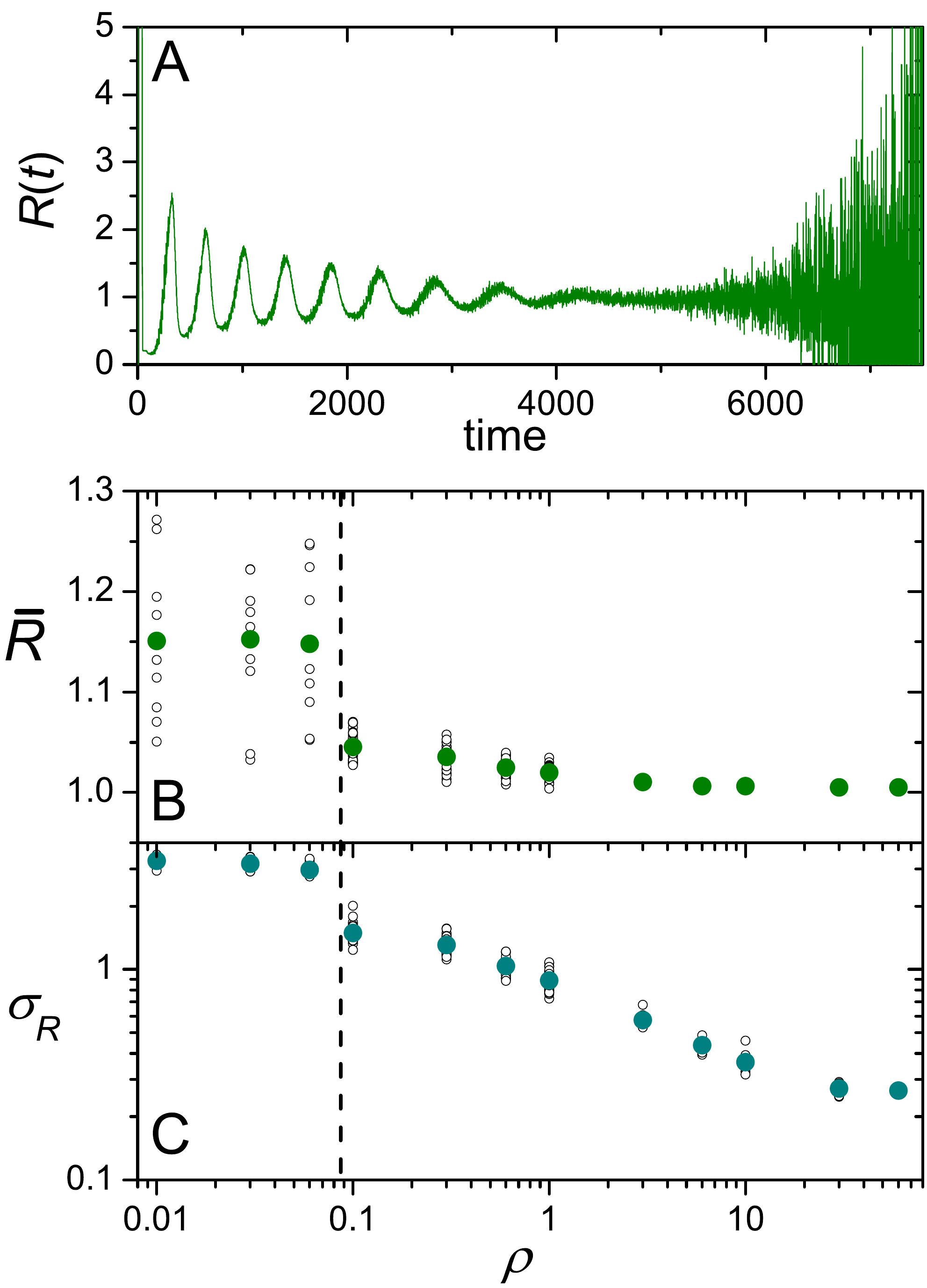}
\caption{Effective reproduction number in the epidemic model, $R(t)$. (A)  Time evolution of  $R(t)$ along a single realization with $\lambda=0.05$, $\omega=0.05$, and $\rho=1$. The effective reproduction number oscillates around $R=1$ and has large fluctuations when the fraction of infected individuals is small, especially when the epidemic is ending. (B) Time average  of the effective reproduction number, $\bar R$,  as a function of $\rho$. Small open symbols correspond to single realizations, and solid symbols stand for the average over realizations for each value of $\rho$.
(C) Standard deviation  of  the effective reproduction number along the dynamics, $\sigma_{R}$, for the realizations shown in (B). Solid symbols are averages over realizations.}
\label{fig:R1}
\end{figure}

The dynamics of $R(t)$ for an illustrative example are displayed in Fig.~\ref{fig:R1}A. Oscillations around $R \approx 1$, which turn out to be synchronous with those of $P(t)$ and $q(t)$, have a large amplitude at short times due to the influence of the initial condition.  At later times, $R(t)$ displays increasingly stronger, seemingly random fluctuations, when the number of infected individuals declines (compare Fig.~\ref{fig:R1}A with Fig.~\ref{fig:ModelDynamics}B). However, the time average of $R(t)$ along the whole dynamics (from $t=1$ to the time of extinction $t_{\rm ext}$),
\begin{equation}
    \bar R = \frac{1}{t_{\rm ext}} \sum_{t=1}^{t_{\rm ext}} R(t),
\end{equation} 
has a definite value which converges to one as $\rho$ increases and infection waves become more numerous. This convergence is shown in panels B and C of Fig.~\ref{fig:R1}. For small $\rho$ (to the left of the dashed vertical line), the infection ends after a single wave, and the value of $\bar R$ varies strongly between different realizations. As more waves develop, on the other hand, $\bar R$ is less disperse and, at the same time, it is closer and closer to one.  

This robust behavior towards $R=1$ does not hold if the population ignores the external risk. If  $\omega=\rho=0$, in fact, there is always a single infection wave that ends once $R(t)$ has dropped below one and the number of new infections is not able to compensate the removal of infected individuals. In this case, the final value of $R$ depends on $\lambda$ (see Fig.~S3). 

\subsection*{Epidemic shapes the collective risk-taking propensity profile}

The heterogeneity of the population manifests itself along the epidemic outbreak. The probability that an individual becomes infected depends on its basal RTP and on its momentary risk exposure. The first infection wave  affects preferentially those individuals with a high value of $p_i^0$. The rapid progress of the outbreak causes a collective inhibitory response, but it still generates a high number of contagions and also affects individuals with RTP well below average. 

\begin{figure}[h!]
\centering
\includegraphics[width=.9\columnwidth]{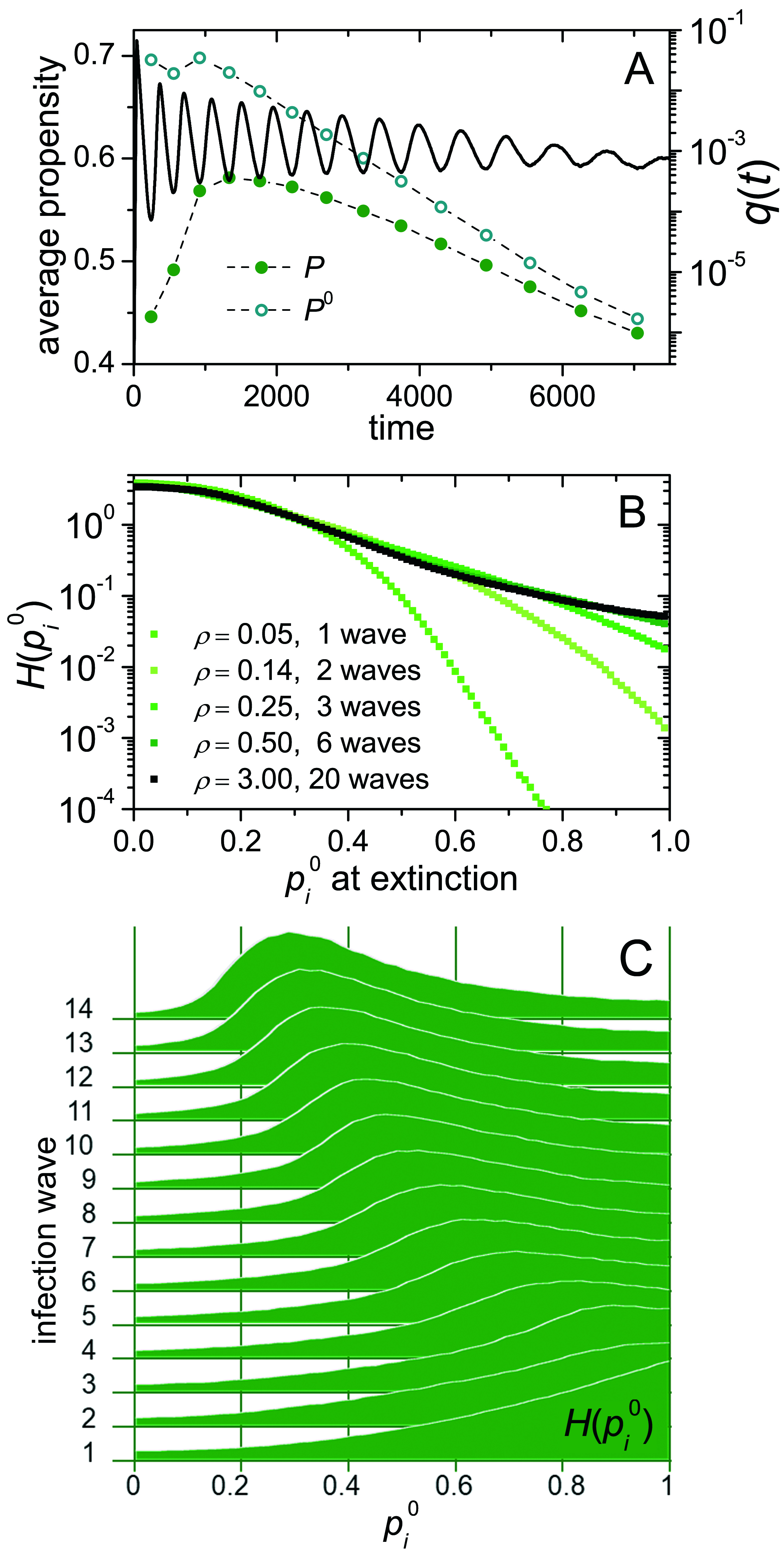}
\caption{(A) The curve shows the evolution of the infected fraction $q(t)$ during a single realization with $\lambda=0.05$, $\omega=0.05$, and $\rho=2$. Full and open symbols respectively stand for the average RTP $P$ and the average basal RTP $P^0$ at the moment of infection, after each infection wave. (B) Normalized histograms of the individual basal RTPs $p_i^0$ over the surviving susceptible population at the time of extinction, for $\lambda=0.05$, $\omega=0.05$, and different values of $\rho$, corresponding to different numbers of infection waves, as indicated in the legend. Data were recorded over $10$ to $50$ realizations for each parameter set. (C) Histograms of  basal RTPs $p_i^0$ for the individuals infected at each wave, from $1$ to $14$.}
\label{fig:RiskProfiles}
\end{figure}

Figure~\ref{fig:RiskProfiles}A shows the average RTP of individuals infected at each wave for an epidemic with many waves. The quantities $P$ and $P_0$ plotted in the figure are, respectively, averages of $p_i(t)$ and $p_i^0$ over all the individuals which, during the interval elapsed between two consecutive minima of $q(t)$, have undergone the transition from susceptible to infected. The momentary RTP $p_i(t)$ taken into account to compute the average was recorded at the moment of contagion. The resulting values of $P$ and $P^0$ are plotted at the time of the minimum just after the corresponding interval.   

Overall, the average basal RTP $P^0$ decreases with each subsequent infection wave. On the other hand, $P$ grows during the first few waves, and then decreases much as $P^0$. The initial growth of $P$ is a direct consequence of the fact that, at the very beginning of the infection, individual RTPs drop abruptly because of the fast growth in the fraction of infected population. Although individuals with higher RTP are more prone to undergo contagion, on the average, the population that becomes infected has a relatively low value of $P$. In successive waves, this effect weakens and the average  RTP of the susceptible population grows, which favors contagion with higher values of $P$. Progressively, however, individuals with high basal RTP are removed from the susceptible population which leads to a collective decrease in the momentary RTP. The average $P$ thus starts decreasing, and eventually mirrors the behavior of $P^0$ (see also Fig.~S4).

The preferential infection of individuals with high RTP leads to a collective process of self-organization, revealed by the fact that the distribution of basal RTPs $H(p_i^0)$ in the remaining  susceptible population acquires a well-defined profile. Figure~\ref{fig:RiskProfiles}B shows the numerical estimation of this distribution at the moment of infection extinction, for epidemics with an increasing number of waves. The selection of individuals with small values of $p_i^0$ in detriment of those with high basal RTP is apparent. It is worth mentioning that, for  $\omega=\rho=0$, $H(p_i^0)$ exhibits an exponential profile which strongly depends on $\lambda$ (see Fig.~S3). 

Figure~\ref{fig:RiskProfiles}C presents histograms of basal RTPs for the individuals that become infected at each successive wave. Note that, in accordance with  the data for $P^0$ in Fig.~\ref{fig:RiskProfiles}C, the maximum of $H(p_i^0)$ monotonically shifts to lower values of the basal RTP.

\section*{Mean-field model}  

The dynamical interplay of epidemiological variables and RTP that gives rise to the infection waves observed in our  stochastic model can be understood in terms of a simplified continuous-time mean-field version, in the form of a system of ordinary differential equations. These read
\begin{eqnarray}
\dot s(t) &=& - P(t)s(t) q(t),\nonumber \\
\dot q(t) &=& \  \ P(t)s(t) q(t)- q(t), \label{sda} \\
\dot P(t) &=&- w P(t)q(t)+ r P(t) [1-q(t)]  \theta [P_0-P(t)],\nonumber \end{eqnarray}
where dots indicate time derivatives. As in the main text, $s(t)$ and $q(t)$ are here the fractions of susceptible and infected individuals, respectively, while $P(t)$ represents the average RTP of the susceptible population. In the first two equations, we recognize an SIR model with contagion rate $P(t)$. The infection frequency of the standard SIR model, namely, a constant coefficient which would multiply the product $s(t)q(t)$ on the right-hand side of the two first equations, has been absorbed by $P(t)$. Hence, in contrast with the stochastic model, $P(t)$ is here not necessarily limited to the interval $(0,1)$. Moreover, a rescaling of time allows us to fix the removal frequency, corresponding to the probability $\lambda$ of the stochastic model,  to unity.  

The first term on the right-hand side of the third of Eqs.~\ref{sda} stands for the decrease of the RTP weighted by the fraction of infected population. Instead of the random threshold which avoids that the individual propensities  reach non-positive values in the stochastic model, in the mean-field version the decrease of the average RTP is proportional to $P(t)$ itself, which insures that $P(t)$ remains always positive. In the second term, which represents the RTP growth, the Heaviside step function $\theta [P_0-P(t)]$ describes the saturation of $P(t)$ to an upper basal average value $P_0$. The frequencies $w$ and $r$ respectively control the rates at which $P(t)$ decreases and increases. They are the continuous-time counterparts to the parameters $\omega$ and $\rho$  of the stochastic model. 

\begin{figure}[t]
\centering
\includegraphics[width=.9\columnwidth]{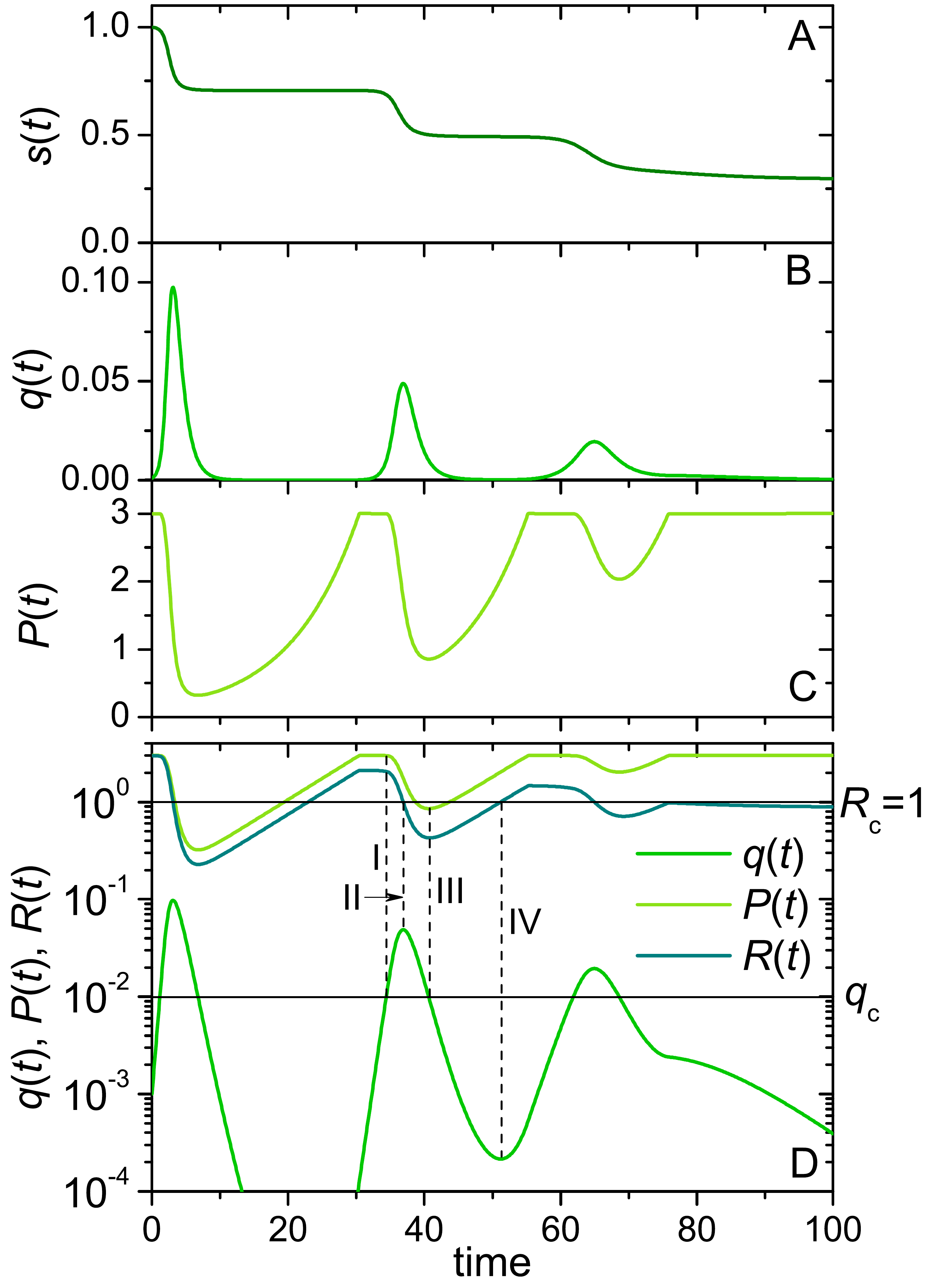}
\caption{Numerical solution to Eqs.~\ref{sda} for (A) the susceptible fraction $s(t)$, (B) the infected  fraction $q(t)$, and (C) the average RTP $P(t)$, with $w=10$, $r=0.1$, and $P_0=3$. The initial condition is $s(0)=0.999$, $q(0)=0.001$, and $P(0)=P_0$. (D) Log-linear plot of  the time evolution of the infected  fraction $q(t)$,  the average RTP $P(t)$, and the effective reproduction number $R(t)=P(t) s(t)$, for the same numerical solution. Horizontal lines show the critical effective reproduction number $R_c=1$ and the critical infected fraction $q_c=r/(w+r)$. Vertical dashed lines with Roman numerals indicate, across the second  infection wave, the events that control the occurrence of oscillations. I: $q(t)$ crosses  the critical value $q_c$ upwards, and $P(t)$ begins to decrease from $P_0$; II: $R(t)$ crosses  $R_c=1$ downwards, and $q(t)$ attains a maximum; III: $q(t)$ crosses  $q_c$ downwards, and $P(t)$ attains a minimum; and IV: $R(t)$ crosses $R_c$ upwards, and $q(c)$ attains a minimum. In this particular example,  after the third infection wave  ($t \approx 75$),  the susceptible fraction is not large enough to allow $R(t)$ to cross $R_c$ again, and the infection proceeds  to extinction without further oscillations.} \label{fig:SI0}
\end{figure}

In the mean-field model, the product $R(t)=P(t) s(t)$ is the epidemiological effective reproduction number. In fact, it is clear from the second of Eqs.~\ref{sda} that the infected population grows ($\dot q>0$) or shrinks ($\dot q<0$) depending on whether the product $P(t)s(t)$ is respectively greater or less than the critical value $R_c=1$. When $R(t)=R_c$, we have $\dot q=0$, and $q(t)$ attains a maximum or a minimum. When $P(t)<P_0$, likewise, there is a critical value of $q(t)$,
\begin{equation}
    q_c= \frac{r}{w+r},
\end{equation}
at which $\dot P=0$. For $q(t)$ greater or less than $q_c$, $P(t)$ respectively decreases or increases. This coaction between the fraction of the infected population and  RTP is the reciprocal mechanism that induces the oscillatory behavior of our system, as we explain in the following. The whole process is illustrated along a few infection waves in Figure \ref{fig:SI0}, where roman numerals indicate the key events in an infection cycle.

Suppose that, as in the stochastic model, the initial state of the population mostly consists of susceptible individuals, with a small infected fraction. We also assume that, initially, the average RTP is fixed at its basal value $P_0$. If the basic reproduction number $R_0=P_0 s(0)$ exceeds one, the infection spreads and $q(t)$ begins growing. At the same time, if $q(0)<q_c$, $P(t)$ would tend to increase but, as it is already at its saturation value $P_0$, it initially remains constant. However, if the susceptible fraction is large enough, $q(t)$ will eventually reach and overcome $q_c$, at which point $P(t)$ begins to decrease. Since $s(t)$ is steadily decreasing as well,  the effective reproduction number $R(t)$ progressively approaches $R_c$ from above. At the moment when $R(t)$ drops below $R_c$, the infected fraction attains a maximum and begins to shrink, while $P(t)$ and $s(t)$ keep decreasing. This continues until $q(t)$ attains $q_c$ again, now from above, and consequently $P(t)$ reaches a minimum and begins to grow. If the susceptible fraction is still large enough, $R(t)$ can overcome $R_c$ again. The infected fraction begins growing and the cycle restarts.   

This explanation emphasizes the fact that the susceptible population is the ``fuel'' that keeps the infection cycles going. Indeed, a sufficiently large number of susceptible individuals  is necessary to make both the infected fraction and the effective reproduction number overcome their respective critical values $q_c$ and $R_c$. Oscillations terminate when the susceptible fraction drops below a level such that either $\dot q$ or $\dot P$, or both, cannot change their sign anymore. From then on, the infection proceeds monotonically towards extinction.  

Although  the final (asymptotic) effective  reproduction number in the mean-field model is generally not equal to one, successive oscillations of decreasing amplitude around its critical value make $R(t)$  become progressively tuned near $R_c$. This is shown in  Figure \ref{fig:SI0} for just three infection waves, but the larger the number of oscillation the closer to one is $R(t)$ expected to end. This behavior is in close agreement with our results for the stochastic model, as illustrated by Figure \ref{fig:R1}A. 

\section*{Discussion}

The temporal variation of the effective reproduction number of an epidemic emerges from the interplay of multiple factors affecting the spread of the disease, such as the progressive depletion of susceptible individuals,  seasonal changes in transmissibility, imposed modifications in the contact patterns among individuals, and non-pharmaceutical control measures, among others \cite{cori:2013}. The added effect of unsupervised responses to the perception of epidemiological risk can also play a main role in restraining epidemic progression. However, the quantitative relevance of the mechanisms that underlie individual risk-taking propensity is not yet fully understood: it is believed that emotions may drive risk perceptions, sometimes more  than factual information \cite{vanBavel:2020}, and that behavioral responses to pandemics are primary shaped by risk attitudes, and not so much by actual incidence or mortality \cite{chan:2020}. Behavioral sciences have much to add to epidemiology, not only to improve public communication and to understand social reactions to information campaigns \cite{vanBavel:2020}, but also to sort out the main determinants of risky attitudes.  
 
Regardless the fundamental drivers behind individual responses to perceived risk, our results specifically hint at the possibility that such responses tune the effective reproduction number around its critical value $R=1$. In our model, single infection waves are characterized by a unique transition from $R >1$ at the beginning of the epidemic to $R<1$ at later stages, as observed in many epidemic outbreaks \cite{thompson:2019}. However, as the number of infection waves increases, the tuning to the critical value becomes more effective. The number of waves increases when the relaxation of the risk-taking propensity toward its basal (highest) value becomes faster, while the average incidence decreases. Though, once the epidemics terminates, the final outcome is only weakly dependent on response parameters, lower incidence levels should permit a more efficient management of the epidemic, for instance, by avoiding saturation of the health system. 

\begin{figure}[tbhp]
\centering
\includegraphics[width=\columnwidth]{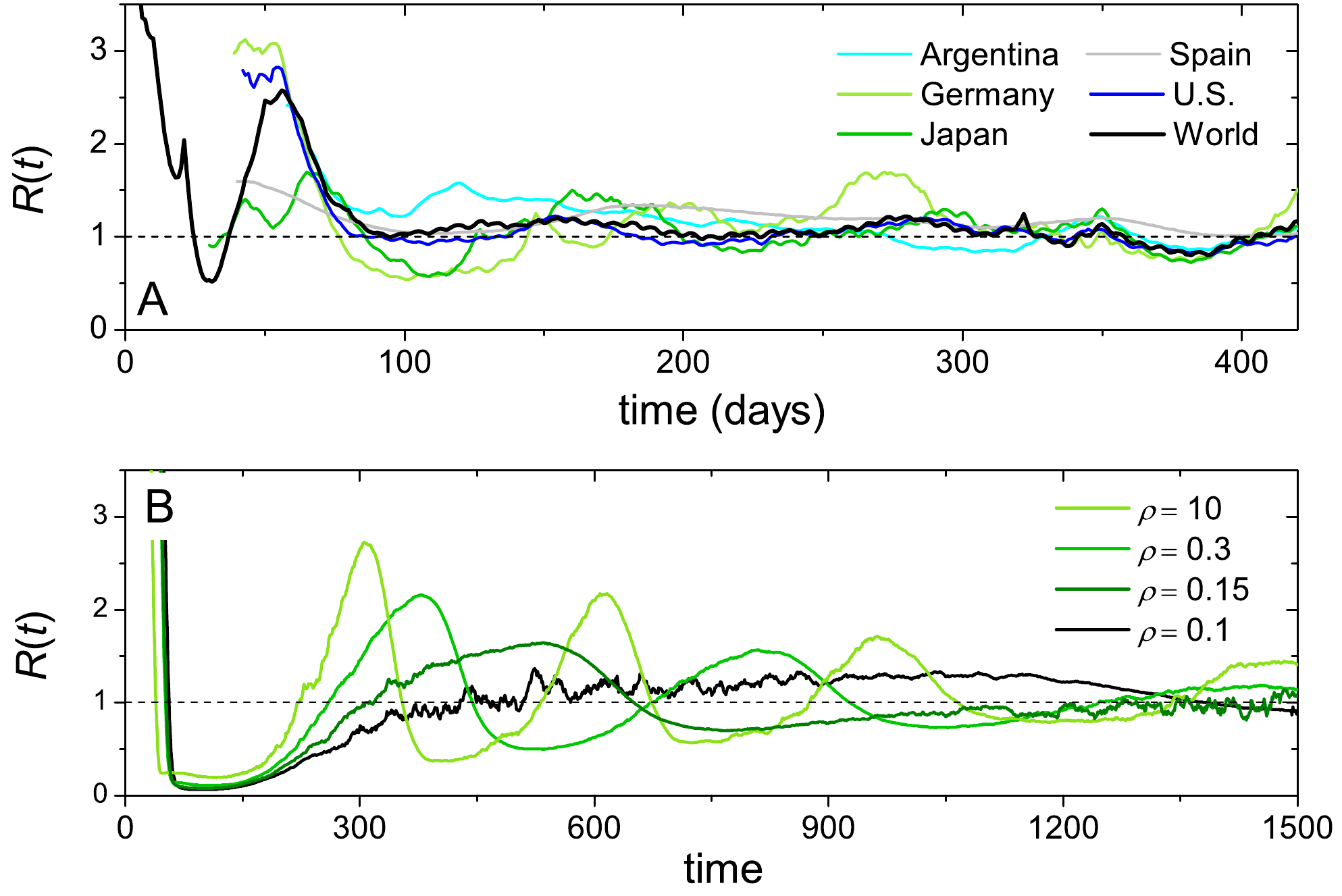}
\caption{Effective reproduction number, $R(t)$. (A) Biweekly running averages of $R(t)$ for several countries and the world, along COVID-19 pandemic since January 23, 2020.  (http://www.globalrt.live/, accessed on March 22, 2021). (B) Time series of $R(t)$ obtained from the stochastic epidemic model for $\lambda=0.05$, $\omega=0.05$, and four values of $\rho$. Each value is a running average over a window of size $\lambda^{-1}$, the average duration of the infection.}
\label{fig:R(t)}
\end{figure}

The emerging stabilization of $R$ around its critical value is in good agreement with empirical observations of the progression of COVID-19 \cite{arroyo:2021,koyama:2021} (see some examples in Fig.~\ref{fig:R(t)}).  Despite local idiosyncrasies, such as the different timing at which institutional control measures are applied and lifted (or even in the absence of such measures), this stabilization holds broad and wide after a few epidemic waves. We hypothesize that it is the unsupervised reaction of individuals to the current epidemic state, after a short learning period and in a way analogous to the inhibition-relaxation response implemented in our model, that overall compensates for other mechanisms affecting epidemic spread. 

It could be  further argued that it is at the point when $R \approx 1$ that non-pharmaceutical measures have to be enforced if the aim is to fully inhibit the epidemic, given that the population may spontaneously enter a relaxation period, thus halting the decrease of $R$. Actually, both self-organized responses and imposed inhibitory measures are typically relaxed as incidence waves are in their decreasing phase. Our model shows that these mutually compensating mechanisms drive the epidemic to an intermediate stage  of low morbidity, that will eventually give rise to a new wave if the fraction of remaining susceptible individuals is high enough. This result is in agreement with other models that identify infection waves as fragile states of transient collective immunity \cite{tkachenko:2021} that degrade as soon as incidence decreases and individuals recover their basal risk exposure levels. 

Epidemic models have a limited predictive ability due to the impossibility of implementing all nuances of the real world, but also to incomplete data that enhances the inherent sensitivity to model parameters \cite{castro:2020}. This nonetheless, models do have an important explanatory ability, by clarifying the individual effect of each involved mechanism. In this sense, our model has to be understood as a proof-of-concept: it shows that the collective response of a population with heterogeneous and time-varying risk propensities suffices to reproduce sustained oscillations and the empirical trend towards $R=1$. Several variations of the model might be worth exploring, including the possible effect of local versus global information, the dependence on different epidemic cues in individual responses, or the role played by other sources of heterogeneity, as in the number of contacts. In this respect, an important open question is whether different models that point at human behavior as the main driver of epidemic dynamics could be synthesized into a small number of generic mechanisms, and if emerging epidemic properties, independent of specific details, do actually exist. At present, such dynamical properties have been ascribed to heterogeneity in individual activity \cite{tkachenko:2021}, to a delayed, awareness-driven population response \cite{weitz:2020}, or to changes in momentary risk-taking propensity, as here suggested, among several other proposals that, however, demand an external regulation of response strategies. 
 
Epidemics have profoundly changed human societies. How they affect social habits, and thus long-term cultural features, is well documented \cite{snowden:2019}. On the evolutionary time scale, epidemics should also impact basal risk-taking propensity, considering that risk aversion emerges by natural selection if reproductive risk is correlated across individuals in a given generation \cite{zhang:2014} --a situation brought about by epidemics. The model  introduced here  may serve as a first quantitative approach to establish a relationship between epidemic characteristics and the strength of selection of risk-aversion profiles. 

\section*{Methods}
\subsection*{Numerical model and algorithm} We simulate the evolution of a population of $N$ individuals where, at each  time step, each individual can be in one of three epidemiological classes: S (susceptible), I (infected), or R (removed). Risk-taking propensity (RTP) is attributed to each individual $i$ in the form of a time-dependent probability $p_i (t)$ varying within the interval $(0,p_i^0)$, where $p_i^0 \in (0,1)$ is the basal RTP. 

The population evolves as follows:
\begin{enumerate}
    \item First, as a response to risk perception, the RTP of susceptible individuals is inhibited, and $p_i(t)$ decreases to $p_i(t) - \omega q(t)$, where $q(t)$ is the fraction of infected population. The positive parameter $\omega$ controls the overall speed of decrease. If $p_i(t)$ drops below $0$, it is reset to $p_i(t) = 0.025 \xi (t)$, with $\xi (t)$ a random number drawn from a uniform distribution in $(0,1)$.
    This resetting to a positive value implements a small but unavoidable probability of contracting the disease. 
    \item Then, with a probability given by the new value of $p_i(t)$, each susceptible individual is exposed to the infection, and
     \begin{enumerate}
         \item with probability $q(t)$,  moves from the susceptible to the infected class;
         \item with the complementary probability, $1-q(t)$, the individual overcomes the threat of contagion. In that case, confidence increases, and  $p_i(t)$ changes to  $p_i(t) + \rho \xi(t)$, with $\xi (t)$ a random number drawn from a uniform distribution in $(0,1)$. This random increase, weighted by the positive parameter $\rho$, accounts for the multiple factors that make $p_i(t)$ relax back to the basal RTP. If, due to this relaxation,  $p_i(t)$ overcomes $p_i^0$,  we reset $p_i(t) = p_i^0$.
     \end{enumerate}
    \item With probability $\lambda$, each infected individual is removed. 
    \item The process terminates when all infected individuals have been removed.
\end{enumerate}

\subsection*{Initial conditions and parameters} At the initial time, the number of susceptible, infected, and removed individuals in the population is $N-1$, $1$, and $0$, respectively. Risk-taking propensities
$p_i(0)=p_i^0$ are drawn from a uniform distribution in $(0,1)$, and $\lambda=0.05$. The resulting average duration of the individual infection is $\lambda^{-1}=20$ time steps.  

\begin{acknowledgments}
The authors acknowledge fruitful discussions with Saúl Ares, Mario Castro, José A. Cuesta and Anxo Sánchez. Research supported through the Spanish Ministerio de Ciencia, Innovaci\'on y Universi\-da\-des-FEDER funds of the European Union, project MiMevo (FIS2017-89773-P, SM). The Spanish MICINN has also funded the ``Severo Ochoa'' Centers of Excellence to CNB, SEV 2017-0712, and the special grant PIE 2020-20E079 (SM) entitled ``Development of protection strategies against SARS-CoV-2''. 
\end{acknowledgments}

\bibliography{bibliography-risk}

\clearpage
\newpage
\onecolumngrid
\renewcommand\thefigure{\thesection S\arabic{figure}}  
\setcounter{figure}{0} 

\appendix

\section*{Supplementary Information}

\begin{figure}[h]
\centering
\includegraphics[width=0.45\linewidth]{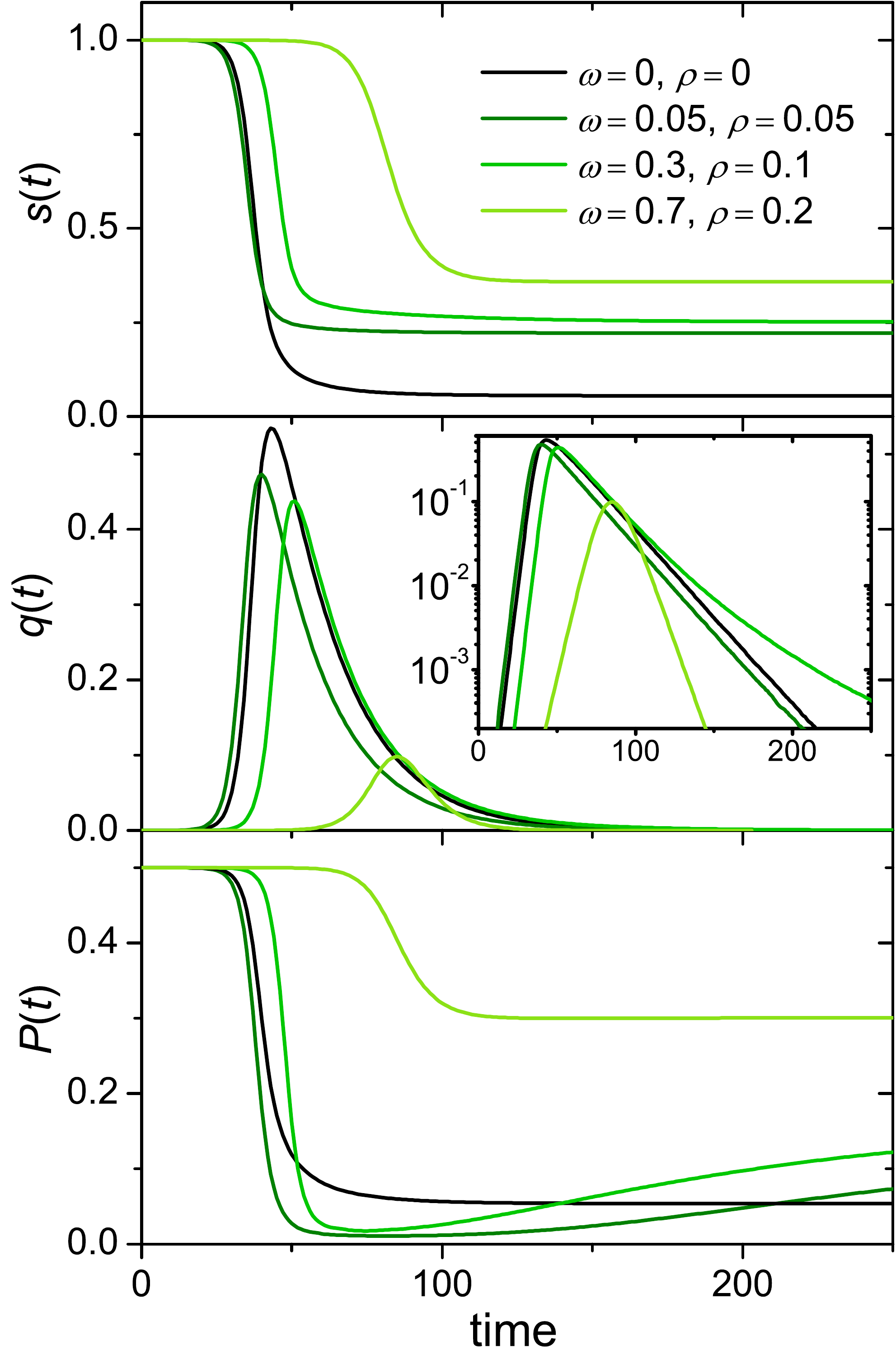}
\caption{Inhibition of infection outbreaks when response parameters are turned on. The figure shows the time evolution of the fraction of susceptible and infected population, $s(t)$ and $q(t)$, and the average risk propensity of susceptible individuals, $P(t)$, comparing the progression of the disease in the absence of population response ($\rho=0$, $\omega=0$) to three examples where the population reacts to risk. In all cases with response, the final impact of the epidemic is diminished. The final fraction of susceptible population increases and the  height of the infection wave decreases as the inhibition rate of risk propensity, measured by the parameter $\omega$, grows.  Meanwhile, $P(t)$ can show non-monotonic evolution for intermediate values of $\omega$, but its final level grows as $\omega$ becomes larger. In the three cases with response shown in the figure, for better comparison with the non-response case, $\rho$ has been chosen in such a way that epidemic terminates after a single infection wave. For each value of $\omega$, the range of $\rho$ where a single wave occurs is limited to small values: further increase in $\rho$ would cause subsequent infection waves.  
}
\label{fig:SI1}
\end{figure}
\clearpage
\begin{figure}[h]
\centering
\includegraphics[width=0.55\linewidth]{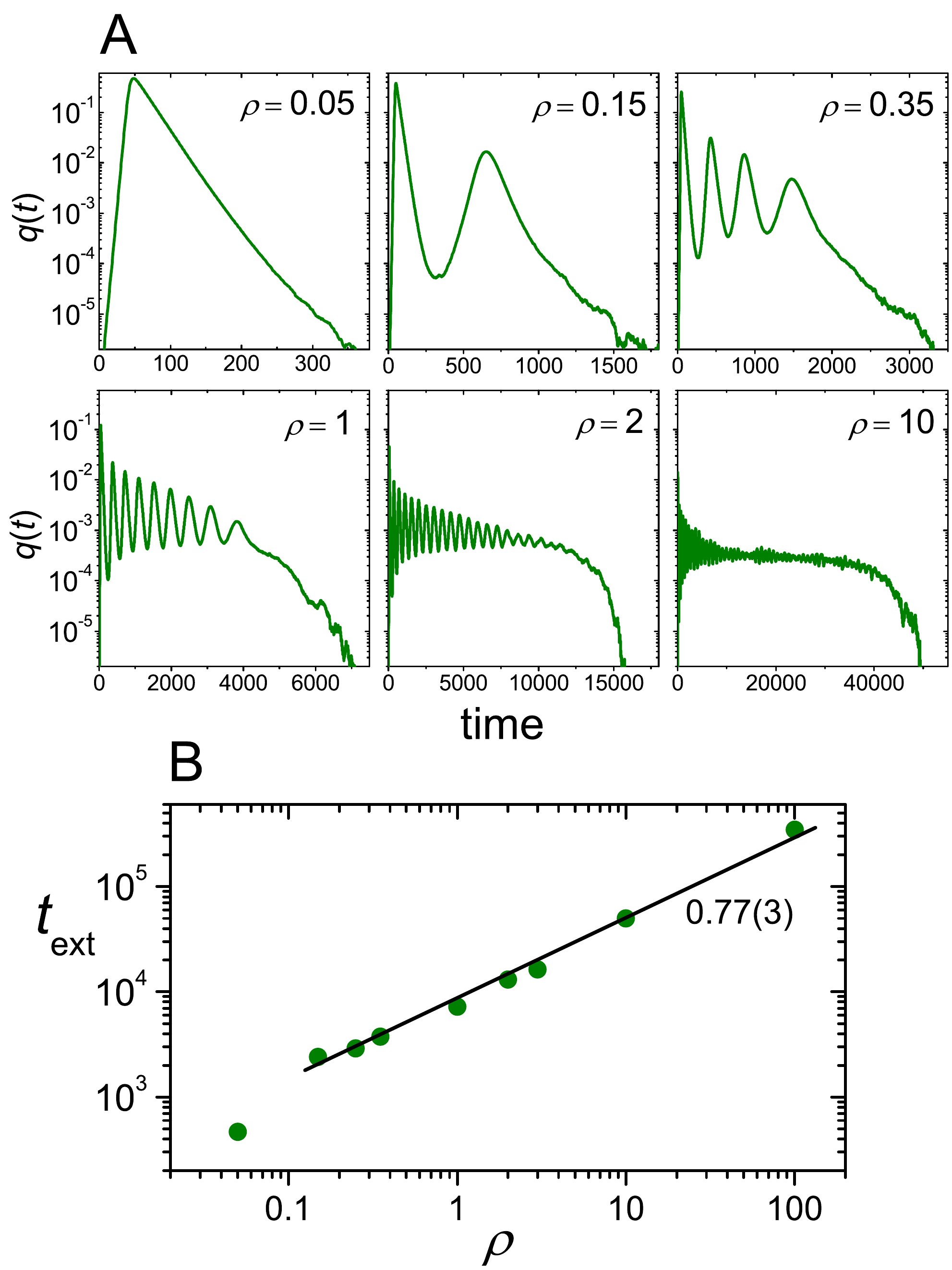}
\caption{(A) Growth in the number of waves as $\rho$ increases, for $\omega=0.05$, $\lambda=0.05$, and $N=10^7$. Panels show the dynamics of the fraction of infected population $q(t)$ for the values of $\rho$ indicated in the legends. For sufficiently large $\rho$, individual waves are no longer seen, as they fuse into a plateau with nearly constant incidence rate decorated with noisy small oscillations. (B) Dependence of the extinction time $t_{\rm ext}$ on $\rho$. The straight line is the best fit to the cases with at least two waves, which yields $t_{\rm ext} \propto \rho^{0.77(3)}$. 
}
\label{fig:SI2}
\end{figure}
\clearpage
\begin{figure}[tbhp]
\includegraphics[width=0.45\columnwidth]{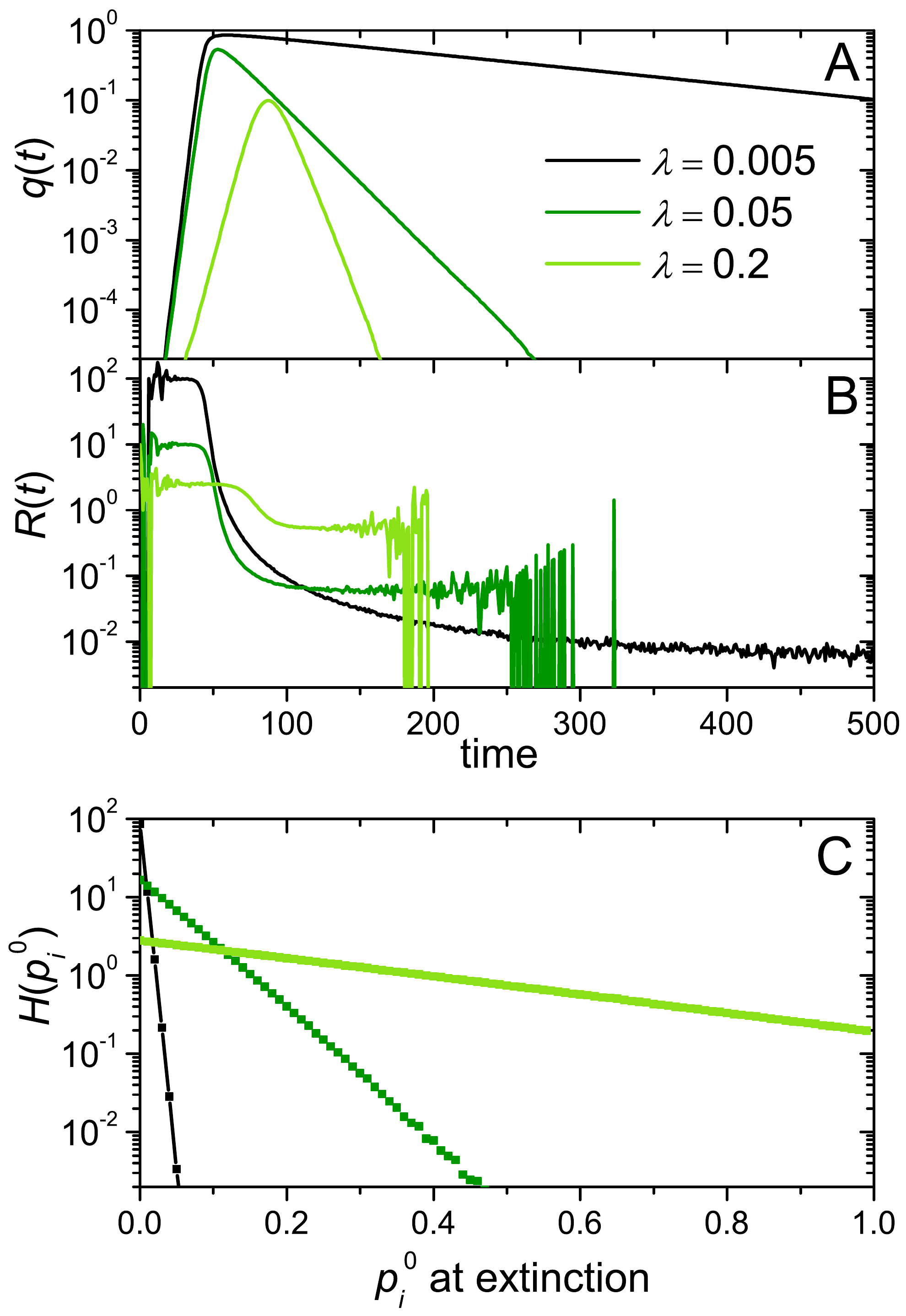}
\caption{Infection dynamics without population response, $\omega=\rho=0$, for various values of $\lambda$, as shown in the legend. (A) Fraction of infected individuals $q(t)$. (B) Effective reproduction number $R(t)$.  (C) Normalized histograms of the individual risk propensities $p_i^0$ over the surviving susceptible population at the time of extinction. In the three cases, the profile is close to an exponential, $H(p_i^0) \propto \exp (-\gamma p_i^0)$, with $\gamma \propto \lambda^{-1}$. 
}
\label{fig:SI3}
\end{figure}
\clearpage
\begin{figure}[tbhp]
\includegraphics[width=0.45\columnwidth]{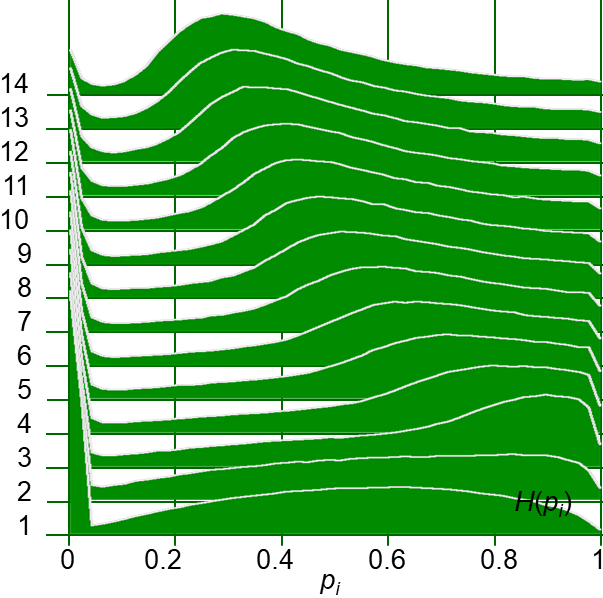}
\caption{Histograms of momentary risk propensities $p_i (t)$ for the individuals infected at each wave in the same simulations as in Fig.~4 of the main text. These distributions have a peak at low values of $p_i$ due to the small but unavoidable probability of getting infected, even under large inhibition of the risk propensity. The peak is particularly high at the first waves due to the larger incidence of the disease, and decreases as infection waves proceed. Besides this peak, $H(p_i)$ attains a maximum at intermediate values of $p_i$ which, in the first few waves, moves to the right and then shifts backward (see Fig.~4B and compare with Fig.~4C).
}
\label{fig:SI4}
\end{figure}

\end{document}